\documentclass[journal]{IEEEtran}
\usepackage{multirow}
\usepackage{array}
\usepackage[ruled,linesnumbered]{algorithm2e}
\usepackage{cite}

\ifCLASSINFOpdf
\usepackage[pdftex]{graphicx}
\else
\fi
\usepackage{makecell}

\usepackage{amsmath}									
\usepackage{amssymb}

\usepackage{listings}

\usepackage{color}

\begin{document}
%
\title{SIGMA : Strengthening IDS with GAN and Metaheuristics Attacks}
%
%
%

\author{Simon Msika, Alejandro Quintero, Foutse Khomh
\thanks{M. Msika was with the Department
of Genie Informatique and Genie Logiciel from Polytechnique Montreal, Montreal,
QC, Canada. e-mail: simon.msika@polymtl.ca}
\thanks{Manuscript received November XX, 2019; revised XXX.}}

\maketitle

\begin{abstract}
An Intrusion Detection System (IDS) is a key cybersecurity tool for network administrators as it identifies malicious traffic and cyberattacks. With the recent successes of machine learning techniques such as deep learning, more and more IDS are now using machine learning algorithms to detect attacks faster. However, these systems lack robustness when facing previously unseen types of attacks. With the increasing number of new attacks, especially against Internet of Things devices, having a robust IDS able to spot unusual and new attacks becomes necessary. \par

This work explores the possibility of leveraging generative adversarial models to improve the robustness of machine learning based IDS. 
More specifically, we 
propose a new method named SIGMA, that leverages adversarial examples to strengthen IDS against new types of attacks. Using Generative Adversarial Networks (GAN) and metaheuristics, SIGMA 
generates adversarial examples, iteratively, and uses it to retrain a machine learning-based IDS, until a convergence of the detection rate 
(i.e. until the detection system is not improving anymore).
A round of improvement consists of a generative phase, in which we use GANs and metaheuristics to generate instances ; an evaluation phase in which we calculate the detection rate of those newly generated attacks ; and a training phase, in which we train the IDS with those attacks.

We have evaluated the SIGMA method for four standard machine learning classification algorithms acting as IDS,
with a combination of GAN and a hybrid local-search and genetic algorithm, to generate new datasets of attacks.
Our results show that SIGMA can successfully generate 
adversarial attacks against different machine learning based IDS. 
Also, using SIGMA, we can improve the performance of an IDS to up to 100\% after as little as two rounds of improvement. 

\end{abstract}

\begin{IEEEkeywords}
IEEE, IEEEtran, Cybersecurity, IDS, Deep Learning, Machine Learning, GAN, Metaheuristics.
\end{IEEEkeywords}

%
\IEEEpeerreviewmaketitle

\section{Introduction}
%
%
%
%
\IEEEPARstart{I}{n} the last few years, the emergence of the Internet of Things (IoT) has led to new cybersecurity challenges. As connected objects now interact with the real world, privacy and security threats mitigation increasingly become a major issue~\cite{IOTchallenges}. With these new entities come the need to protect them from cyberattacks and similar intrusions.
For instance, in 2016, the Mirai botnet \cite{Mirai} infected more than 600,000 Internet of Things devices from which were conducted massive Distributed Denial of Service (DDOS) attacks against several network companies all over the world. \par
Intrusion Detection Systems (IDS) are an essential tool for IoT system administrators : detecting a cyberattack is the first step to guarantee the privacy and security of users.\par
But IoT also means a huge increase of internet traffic to analyze, and therefore the need to develop efficient, fast and robust algorithms to detect cyberattacks in this sensitive environment. Recently, machine learning models have shown astonishing performances in retrieving patterns from large volumes of data, 
in a very short amount of time. This success lead to their wide adoption in IDS~\cite{MLforIDS}. 
However, as recent works on adversarial models have shown~\cite{GAN}, machine learning algorithms, in particular deep learning tend to be fragile to adversarial examples. Making IDS which are based on such models vulnerable to adversarial attacks. \par
Using Generative Adversarial Networks (GAN) \cite{GAN} an attacker can generate adversarial requests (i.e., attacks) that share the characteristics of requests that are considered to be genuine by the IDS. 
%
Although these GANs represent formidable weapons for attackers, as they can deceive most IDS into classifying attacks as benign traffic, they also provide an opportunity to preemptively strengthen intrusion detection systems against new attacks. By exposing an IDS to generated attacks as a preventive measure, we can prepare for new malicious behaviors.\par 

In this paper, we propose a method to strengthen IDS against generated adversarial attacks, called \textbf{SIGMA}, which stands for \textbf{S}trengthening \textbf{I}DS with \textbf{G}AN and \textbf{M}etaheuristics \textbf{A}ttacks. The method consists in the iterative generation of attack datasets using adversarial learning and metaheuristics algorithms. The generated datasets are then used to retrain IDSs; i.e., teaching them to cope with patterns of attacks similar to those contained in our generated datasets. We repeat the retraining process 
until the detection rate of the IDS on generated attacks converges, meaning the detection rate is not improving anymore. We stop the algorithm after 3 runs without a detection rate improvement. 

We evaluated SIGMA on IDSs based on four different classification algorithms : Neural Network, Random Forest, Support Vector Machine, and Naive Bayes Classifier. Each IDS was composed of two parts : a discriminator, trained to detect generated attacks, and an attack-classifier trained on the original 
dataset,  to classify benign traffic and attacks. 
We trained 
a GAN and ran a local-search and genetic algorithm hybrid \cite{Hybrid} to generate attacks against our IDSs. We compare the results of our model trained with both GAN and metaheuristics generated instances, with a model trained only with GAN generated instances over time and another model trained only with metaheuristics generated instances.  \par
Results show that for IDS consisting of a Neural Network or a Random Forest algorithm, the SIGMA method allowed for a detection rate of 100\% of generated attacks two to three times faster than the model strengthened only with the GAN generated attacks. However, models trained only with the instances created using metaheuristics search were almost completely unable to detect GAN generated attacks, suggesting that 
metaheuristics alone are not sufficient to increase the robustness of the studied IDS. 


\textbf{The remainder of this paper is organized as follows.} Section~\ref{back}  
provides an overview of the technologies used in our model. We discuss the related literature in 
Section~\ref{relatedWork}. Section~\ref{SIGMA} presents our strengthening method to increase the robustness of IDS against generated adversarial attacks (i.e., SIGMA). 
Section~\ref{eval} describes the approach followed to evaluate SIGMA, while Section~\ref{results} discusses the obtained results. 
Section~\ref{threats} discusses threats to the validity of our study and Section~\ref{implications} presents some implications of our work. Finally, Section~\ref{conclusion} concludes the paper, summarising our findings 
along with some recommendations for future work.

\section{Background}\label{back}

This section provides background information about Generative Adversarial Networks and metaheuristics used in this paper.

\subsection{Generative Adversarial Networks}
Generative Adversarial Networks are a class of unsupervised machine learning 
algorithm. They are composed of 
two neural networks : a generator $G$ and a discriminator $D$. \par 
Considering a dataset, the generator generates new data instances similar to the ones in the dataset. The discriminator, on the other hand, evaluates the data authenticity, i.e., whether or not the data it reviewed belongs to the actual dataset. The goal of the generator is to generate data labeled as genuine by the discriminator. \par
The generator takes random numbers as input (referred to as random noise), and returns a data instance.
\par
With $x$ as input of the discriminator $D$, we represent the probability that $x$ is an attack generated by $G$ as D($x$). Therefore, D($x$) is equal to zero when $x$ is considered as an authentic data from the dataset, and equal to one when $x$ is labeled as generated data, or fake. \par
With $z$ as random noise, we represent the instance generated by the generator network $G$ as G($z$). \par
The generator $G$ is trained to maximize the function $1 - D(G($z$))$. \par

\begin{figure}[ht]
    \begin{center}
    \hspace*{-0.05in}
        \includegraphics[width=1.1\linewidth]{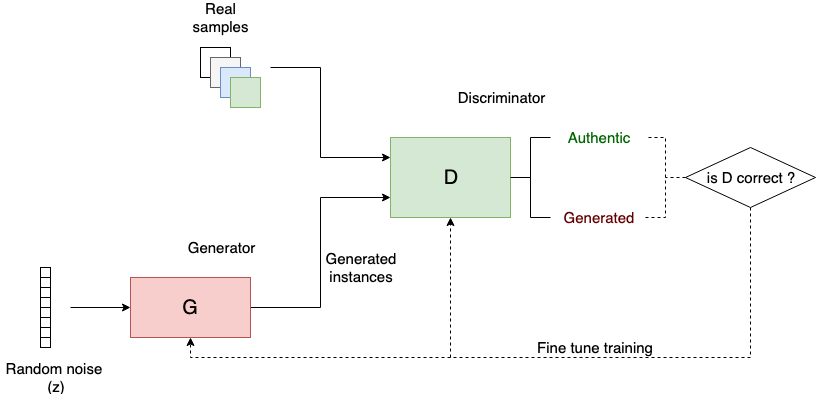}
    \end{center}
	
  	\caption{Diagram of a Generative Adversarial Network (GAN).}
  	\label{fig:GAN}
\end{figure}

As shown on Fig.~\ref{fig:GAN}, the Generator and the Discriminator are trained simultaneously, therefore being on a constant battle throughout the training process. 

This algorithm has rapidly grown in popularity thanks to its performance in image generation \cite{photorealistGAN}. It can generate 
realistic examples, and has a better performance than Deep Belief Networks or Boltzmann Machines \cite{GAN}. \par

GAN are also notably used to disrupt trained classifiers \cite{disruptiveGAN}: slight controlled modifications to the original input leads to misclassification. This has been extensively applied to images classification, due to its impressive results in this field: for instance, small visible changes made to ``Stop'' traffic signs tricked autonomous cars into misclassifying them into speed limit signs \cite{AIsigns}. 

In this paper, we will use this ability of GANs to disrupt trained classifier by training them to generate attacks able to bypass the detection algorithm, \textit{i.e.,} attacks classified as benign traffic by our IDS. Adding subtle modifications to the features of existing attacks could in fact lead to misclassification.

\subsection{Metaheuristics}

A metaheuristic is an algorithm used to find, generate, or select a heuristic (i.e., a partial search algorithm) that can provide a sufficiently good solution to an optimization problem with incomplete information. This type of algorithm is usually employed to solve computationally hard problems for which regular optimization would be too costly. Even if they do not guarantee finding the optimal solution for the problem, they usually provide good results, often close to the optimal solution~\cite{HGAspeed}.  
A metaheuristic approach could be either single-solution based or population based. \par

A single-solution based approach could be local optimization: we randomly initialize a solution and explore the $neighbourhood$ of the solution by applying local changes to the current solution. The search continues until a solution meeting the initial stopping criteria is found or a time bound is elapsed. Local optimization could be very effective in case the criterion to maximize only has a single optimum. 
In other cases, the local search algorithm can converge to a local optimum, therefore not giving the best possible solution. \par
The hill climbing algorithm is an example of local optimization algorithm. This algorithm is an iterative algorithm that tries to improve a solution by making an incremental change to it. If the change produced a better solution, then it becomes the new solution, and another incremental change is made to this new solution. This algorithm runs until there is no further improvement possible. \par

Genetic algorithm~\cite{Genetic} is an example of a population based metaheuristic. It is inspired by the process of natural selection. It starts with a population of solutions, where each solution is randomly generated. The population then evolves until the stopping criteria is met or until a certain number of generations is reached. From this pool of solutions, we select the best solutions (selection) and recombine them into a new population of solutions (crossover). We then apply random mutations to this population, in order to have a diverse population of solutions and possibly exploring other parts of the solution space that were not explored yet. As a global search algorithm, the genetic algorithm metaheuristic is more likely to find global optima for multimodal functions but it is slower at converging\cite{SGA}.  \par

Since the genetic algorithm is rather slow to converge, it is possible to combine those two approaches (local optimization and a population based solution) to have a faster convergence. We then refer to this method as an hybrid algorithm \cite{Hybrid}. It consists of a slight modification of the genetic algorithm to incorporate a local optimization element: after the selection process, we optimize each solution of the population with the local search algorithm. This leads to overall better results, since the local search can only improve a solution, and could mean a faster convergence~\cite{HGAspeed}.

The overall processing of the hybrid algorithm
is exposed in algorithm \ref{alg:hybrid}.

\SetAlFnt{\small}

\begin{algorithm}
\SetAlgoLined
\KwIn{Instance (I), size of population ($\alpha$), selection rate($\beta$), mutation rate (m), number of iteration (nb\_it), Local search algorithm (local\_search)}
\KwOut{Population of solutions to I}
\tcc{Initialization}
\tcc{Generate $\alpha$ random solutions to I}
solutions = generate\_random($\alpha$)\;

\For{i = 1 to nb\_it}{
 \tcc{Local Search}
 solutions = local\_search(solutions)\;
 \tcc{Selection}
 n = $\alpha * \beta$ \;
 pop = select\_best(n, solutions) \;
 \tcc{Crossover}
 p = $(\alpha - n)$\;
 \For{j = 1 to p}{
   randomly select $Sol_A$ and $Sol_B$ from solutions\;
   generate $X_{AB}$ by mixing $Sol_A$ and $Sol_B$\;
   save $X_{AB}$ to offsprings\;
  }
  
 \tcc{Mutation}
 \For{child in offsprings}{
  mutate child with probability m\;
 }
 solutions = pop + offsprings
}
\Return{solutions}
\caption{Pseudocode of the hybrid algorithm}
\label{alg:hybrid}
\end{algorithm}

This hybrid method has often been used to solve complex problems with good results~\cite{HGAspeed}.

In the intrusion detection domain, such algorithms can be used to generate attacks that the intrusion detection system is unable to detect. In this case, a solution would be an actual attack, and all operations (cross-over, mutation, ...) would be modifications of the attacks features.

\section{Related work}\label{relatedWork}

Over the last two decades, researchers have built several intrusion detection datasets by extracting different network features from real networks during cyberattacks. \cite{KDD99} \cite{newdataset}
Different machine learning algorithms have been explored to build IDS: 
from a simple feed-forward neural network, to Extreme Machine Learning \cite{ELM}, to complex Recurrent Neural Networks \cite{RNNIDS}. Studies show that 
even simple algorithms, such as a Support Vector Machine or J48 decision trees, could lead to good detection results, with 95\% accuracy for the SVM and more than 97\% accuracy for the decision tree~\cite{NSLSVM}. 
Those algorithms could be used in practice by smart objects as intrusion detection systems. In fact, they don't require as much energy as complex Deep Learning models, which is an important factor to consider with such resource-constrained environment \cite{RNNIDS}. 

Nonetheless, these machine learning methods suffer from a severe flaw as an IDS: they are totally vulnerable to new types of attacks. Successful attacks can lead to terrible consequences: economic loss, important privacy issues for smart objects users, etc. 
Moreover, it is now possible to automatically build new attacks against which those systems will be utterly useless, thanks to metaheuristics and Generative Networks, for example. 

The use of metaheuristics for attack generation has been explored by Jan et al.~\cite{Metaheuristic}. In this work, Hill Climbing and Genetic algorithms are used to 
generate malicious XML injections. 
These generated attacks were used for testing purposes but demonstrated the possibility to automatically create attacks using metaheuristics techniques; in particular, the genetic algorithm managed to create a wide variety of attacks evading the web application sanity check more than 95\% of the time. 


Hu et al.~\cite{MalwareGAN} leveraged generative adversarial networks for 
malware detection, by considering the detection algorithm as a black-box (as would an attacker). The attacker does not know the internal structure of the detection model, but only knows the detection result of the detection model under attack. 
Even without having any information on the detection system, this approach led to very impressive results ; GANs deceiving the malware detection algorithm almost every time. \par

Furthermore, recent work has shown that it is possible to generate adversarial examples with intrusion detection datasets. In particular, Lin et al.~\cite{IDSGAN} used a Wasserstein GAN \cite{Wasserstein} to generate adversarial attacks against different classifiers considered as black-box algorithms by the attacker, trained with the NSL-KDD \cite{NSLSVM} dataset. The GAN was able to mislead several classification algorithms into classifying generated attacks as benign traffic. Nonetheless, the NSL-KDD dataset is now 10 years old and its relevance is then questionable. Moreover, researchers have pointed out several problems with NSL-KDD \cite{KDDcritique}, e.g., the lack of Remote to Local and User to Root attacks, as well as the lack of more recent type of attacks~\cite{benchmark}. \par

These last few years, some progress has been made on protecting Intrusion Detection Systems against generated adversarial attacks. Generative models are a double-edged sword, as they can be preemptively used to train the detection model as well. \par
Cordy et al.~\cite{searchbased} created increasingly resilient defense strategies to detect training attacks against a clustering-based IDS. The IDS was improved by simultaneously searching for attacks against the IDS and constantly improving the defense strategy : two genetic algorithms (one for creating attacks, the other to elaborate defense strategies) were used. 
Their resulting system detected 98\% of the generated attacks, whereas the attack generation process systematically found a way to deceive the IDS without defense strategy. This promising result suggests that 
metaheuristics can be successfully used to preemptively strengthen an IDS against generated attacks. 
Nonetheless, this work does not provide any insight regarding the vulnerability of this strategy to other types of generated attacks (GAN generated instances for example). \par
The work of Ferdowsi and Saad~ \cite{DistributedGAN} presents an approach to deploy a distributed intrusion detection architecture capable of detecting adversarial generated attacks. In this work, GANs were trained to generate adversarial attacks, and were then used to train a discriminator, which determined whether the current internet traffic was benign or an attack. However, this system might be susceptible to iterative generated attacks: once the discriminator is trained, it may still be possible to find ways to generate instances able to bypass the detection system. 
An IDS resilient to generated iterative attacks has not yet been explored in the intrusion detection domain.

\section{SIGMA: An approach to improve the robustness of IDSs}\label{SIGMA}

In order to increase the robustness of IDS, 
we propose the following SIGMA method. \par

We take as input a Machine Learning based Intrusion Detection System, and a dataset consisting of attacks and benign traffic.
We iteratively generate attacks with two different methods to train the IDS. \newline
Each training iteration is designated by its number. We note $Score_{i}$ the detection rate of generated attacks by the IDS at iteration number $i$, meaning : \par
$Score_i = \frac{number\: of\: detected \:generated\: attacks}{total\: number\: of\: generated\: attacks}$ \newline

We consider that the generated attacks detection rate of our IDS has converged if :
For $\epsilon > 0$, there exist an iteration number $N$ such that for all iterations $i$ after $N$, we have : \par
$|Score_{i} - Score_{i+1}| < \epsilon $ \newline

The SIGMA method instructions are as follows : \par
While the generated attacks detection rate of our IDS has not converged : 
\begin{itemize}

    \item Step 1 : We train a GAN to generate adversarial attacks against the IDS, considered as a black-box. The goal for this algorithm is to generate attacks deceiving the intrusion detection system. Considering the same notation as in Section II.B, the function to maximize for the generator is $1 - D(G(z))$ where $G(z)$ is the generated attack, $D(x)$ is the probability (computed by the IDS) that $x$ is an attack : the IDS plays the role of discriminator.\par
    At each iteration, the generative algorithm generates new attacks to fool the Intrusion Detection System.
    \item Step 2 : We use the trained GAN to generate attacks against the IDS. We evaluate the score of the detection system for these generated attacks. If the score has not improved for 3 consecutive rounds, we stop the algorithm.
    \item Step 3 : We run a Search-based method in order to search for other possible attacks deceiving the IDS that the GAN might have missed. \par
    The function to maximize for this generative algorithm is : $1 - D(sol)$ where $sol$ is the solution generated by the Search-based algorithm, and $D(x)$ is the probability (computed by the IDS) that $x$ is an attack.
    \item Step 4 : We use the Search-based method to generate attacks against the IDS. We then train the IDS with the generated instances from both algorithms (i.e., GAN and metaheuristics) and with data from the original dataset. Exposing its classifier to real data and generated attacks prevents it from overfitting to generated instances and losing accuracy on other type of traffic.

\end{itemize}

The overall proceedings is illustrated in the algorithm \ref{alg:sigma}. 

\SetAlFnt{\small}
\begin{algorithm}
\SetAlgoLined
\KwIn{IDS to improve (IDS), training set (train\_set), }
\KwOut{Improved IDS}
 converged = False\;
 counter = 0\;
 previous\_score = 0\;
 \BlankLine
 \While{converged = False}{
  \tcc{Step 1 : GAN training}
  generator = GAN.train(IDS, train\_set))\;
  \BlankLine
  
  \tcc{Step 2 : Attack generation and evaluation}
  GAN\_attacks = generator(noise)\;
  predict = IDS(GAN\_attacks)\;
  score = $\frac{nb\_attacks(predict)}{length(GAN\_attacks)}$\;
  \BlankLine
  \eIf{score $\leq$ previous\_score}{
   counter = counter + 1\;
   }{
   counter = 0\;
   previous\_score = score\;
  }
  \BlankLine
  \tcc{If the score has not improved after 3 rounds, we stop the algorithm}
  \If{counter = 3}{
  converged = True\;
  break\;
  }
  \BlankLine
  \tcc{Step 3 : Search-based method}
  search\_based.run(IDS)\;
  SB\_attacks = search\_based.generate()\;
  \BlankLine
  
  \tcc{Step 4 : IDS Training}
  
  IDS.train(GAN\_attacks)\;
  IDS.train(SB\_attacks)\;
  IDS.train(train\_set)\;
  
 }
 \Return{IDS}
 \caption{Pseudocode of the SIGMA process}
 \label{alg:sigma}
\end{algorithm}

By combining attacks from both the Machine Learning and the Metaheuristics methods, we expect to explore a larger 
solution space since the two techniques are significantly different; we expect the generated attacks to be widely distinct. Being confronted with a large sample of diverse attacks, an IDS is likely to gain in robustness. 
\par

\section{Evaluation of SIGMA}\label{eval} 
In this section, we evaluate the effectiveness of SIGMA at improving the effectiveness of an IDS. The quality focus is the improvement of the attack detection rate, through iterative reinforcement using GANs and metaheuristics. The perspective
is that of researchers interested in developing efficient IDS, and practitioners interested in improving the robustness of their IDS. The context consists of the CICIDS2017 benchmark dataset \cite{newdataset}, containing 11 types of networks attacks, and four machine learning-based IDS (i.e., a 3-layers Neural Network, a Random Forest, A Support Vector Machine (SVM), and A Naive Bayes Classifier).
In the following, we provide detailed information about the CICIDS2017 benchmark dataset and the implementation of SIGMA using the four selected machine learning-based IDS.

\subsection{Dataset}
The CICIDS2017 benchmark dataset \cite{newdataset} 
consists of more than 80 network flow features (flow duration, destination port, ...). Table \ref{table:IDSMetrics} provides a summary of those characteristics.
This recent intrusion detection dataset contains 11 types of attacks along with benign traffic. Each entry of the dataset consists of more than 80 columns (namely the extracted network flow features) and is labeled as one of those 11 types of attacks or as benign traffic. 
We grouped the 11 different attacks into four different groups as shown in Table \ref{table:CICIDS}, building four different balanced binary datasets (Attack, Benign), to counterbalance the unbalanced number of attacks per type. 

\begin{center}
\begin{table}[ht]
\caption{Some Network features used by CICIDS 2017.}
\begin{tabular}{ |l|l|} 
Feature name & Description\\
\hline
fl\_dur & Flow duration \\
tot\_fw\_pk & Total packets in forward direction \\
tot\_bw\_pk &Total packets in backward direction \\
fl\_pkt\_s &Number of packets transferred per second \\
ack\_cnt &Number of packets with ACK \\
pkt\_size\_avg& Average size of packet \\
idl\_avg & Mean time a flow was idle\\
\hline
\end{tabular}
\label{table:IDSMetrics}
\end{table}
\end{center}

\begin{center}
\begin{table}[ht]
\caption{Attacks labels and distribution in the CICIDS2017 dataset.}
\begin{tabular}{ l|c|l} 

Attack group & \makecell{Number of\\ attacks} & \makecell{Types of attack} \\
\hline

\multirow{4}{8em}{Denial of Service} & \multirow{4}{3em}{252661} &  DOS Hulk \\ 
& & DOS GoldenEye    \\ 
& & DOS Slowloris   \\ 
& & DOS Slowhttptest  \\
\hline
Distributed DOS & 128027 & DDOS  \\ 
\hline
\multirow{5}{8em}{Bruteforce} &
\multirow{5}{3em}{15342} &

FTP-Patator  \\ 
& & SSH-Patator  \\ 
& & Bruteforce  \\ 
& & Portscan  \\ 
& & Botnet  \\ 
\hline
\multirow{4}{8em}{Infiltration} & \multirow{4}{3em}{720} &
SQL Injection  \\ 
& & XSS  \\ 
& & Heartbleed  \\ 
& & Infiltration  \\ 
\hline

\end{tabular}
\label{table:CICIDS}
\end{table}
\end{center}

We first deleted the constant columns of the dataset, as they don't provide any useful information for classification.  
Data now consists of 71 columns, 70 of them being network flow features, and the last one being the label ($i.e.,$ 0 if it is benign traffic, 1 if it is an attack). \par
Then, since the values of each feature throughout the data widely varies, each column was normalized to have values between 0 and 1. Feature scaling allows for much faster convergence for neural networks.
\par
We normalized data by applying the min-max normalization, namely :\par
$c'_i = \frac{c - c_{min}}{c_{max}}$.\par
Where : 
\begin{itemize}
    \item $c_i$ is the column from the original dataset.
    \item $c'_i$ is the normalized column.
    \item $c_{min}$ is the minimum value of the column.
    \item $c_{max}$ is the maximum value of the column.
\end{itemize}
\par
Each dataset was split into a training set and a test set, respectively representing 90\% and 10\% of the overall dataset.

\subsection{Implementation of SIGMA}

\textbf{Step 1 : GAN training}

\begin{figure}[ht]
    \begin{center}
        \includegraphics[width=\linewidth]{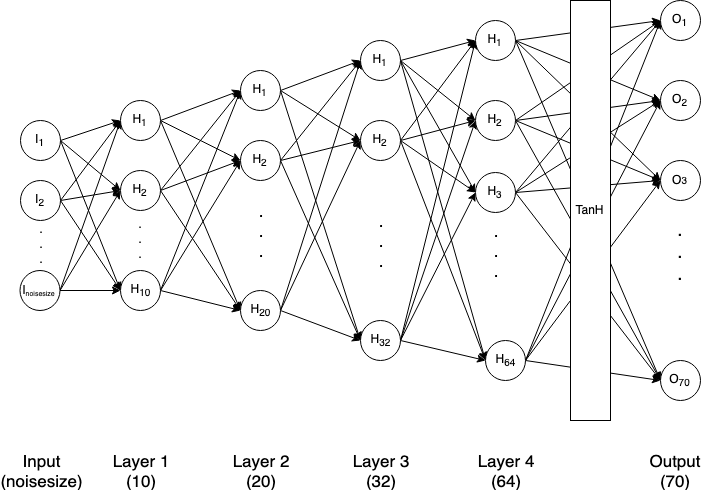}
    \end{center}
	
  	\caption{Architecture of the Generator. Below each layer is shown its dimension. As our GAN takes noise as input, we can make the dimension of this noise vary to change the architecture of the network. The dimension of the output is 70, the number of network features of a dataset entry.}
  	\label{fig:archi}
\end{figure}

We chose to implement SIGMA with a 4-layers Wasserstein GAN. The architecture of the GAN is detailed on Fig. \ref{fig:archi}. The dimensions of hidden layers were chosen experimentally, being the ones with the best results.

As mentioned in Section II.B., the Wasserstein GAN takes random numbers (or random noise) as input to generate attacks. We refer to the number of random numbers as the random noise size. \par

The goal of this generator is to generate attacks able to deceive the IDS. To ensure that the output of the generative algorithm is indeed an attack, we keep the functional features of an original attack. \par
\begin{figure}[ht]
    \begin{center}
        \includegraphics[width=\linewidth]{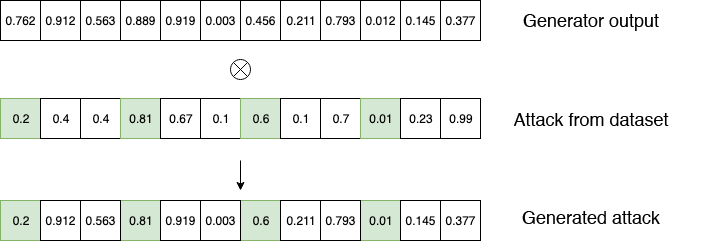}
    \end{center}
	
  	\caption{Diagram of the generative algorithm's process. In green, the functional features of the attack.}
  	\label{fig:expgen}
\end{figure}

Since every feature of our data has been normalized, each feature is represented as a number between 0 and 1. As shown on Fig. \ref{fig:expgen}, we keep the functional features of real attacks for our generated attacks.

The functional features per attack type were identified by a statistical analysis of the datasets, with the help of the analysis conducted by the creators of the dataset\cite{newdataset}. 
They are presented in Table \ref{table:features}.

\begin{center}
\begin{table}[ht]
\caption{Functional features per attack type. Those features are not going to be modified by the generator.}
\begin{tabular}{ | m{2cm} | m{5cm}|} 
\hline

Attack group & Functional features \\
\hline
DOS & Flow Duration, Active Mean, Average Packet Size, \par Packet Length Std, Flow IAT Mean, PSH Flag Count, Idle Max\\
\hline
DDOS & Flow Duration,
 Bwd Packet Length Std,
 Average Packet Size,
 Packet Length Std,
 Flow IAT Std,
 ACK Flag Count\\
 \hline
Bruteforce & PSH Flag Count, 
Flow Duration,
 Total Length of Fwd Packets,
 Init Win bytes forward,
 Packet Length Std,
 Subflow Fwd Bytes, Fwd PSH Flags\\
 \hline
Infiltration & Subflow Fwd Bytes,
 Total Length of Fwd Packets,
Flow Duration,
 Idle Mean,
 Active Mean,
 Init Win bytes backward,
 PSH Flag Count\\
\hline
\end{tabular}
\label{table:features}

\end{table}
\end{center}

With the aim to have a Generative Adversarial Networks with the best performance and therefore explore the largest attack space possible, we trained several Generators with different sizes of noise as input. Furthermore, since there is unpredictability in the training of Generators due to the randomized weights initialization, we trained the Generators several times. We then select the GAN with the best performance among those, \textit{i.e.,} the most able to deceive the IDS. \newline

The process followed to train the GAN is presented in algorithm \ref{alg:GAN}.

\SetAlFnt{\small}

\begin{algorithm}
\SetAlgoLined
\KwIn{IDS (IDS), training\_set (train\_set), Maximum noise size (max\_noise\_size), Number of training epochs (nb\_epoch)}
\KwOut{Trained GAN}
\tcc{Initialization}
best\_score = 1.0\;
noise = 1\;

\For{attempt = 1 to 5}{
 \For{noise\_size = 1 to max\_noise\_size}{
  \tcc{We construct a GAN with the corresponding noise size as input}
  GAN = Generator(noise\_size)\;
  \BlankLine
  
  \For{epoch = 1 to nb\_epoch}{
   \For{(batch, labels) in train\_set}{
    \tcc{First select the attacks from the training set}
    is\_attack = non\_zero(labels)\;
    attacks = select(batch, is\_attack)\;
    
    \BlankLine
    
    \tcc{Then generate attacks}
    z = random\_noise(noise\_size)\;
    generated\_attacks = GAN(attacks,z)\;
    
    \tcc{Backpropagation}
    loss = mean(IDS(generated\_attacks))\;
    loss.backward()\;
    optimizer.step()\;
    
    \If{loss $\leq$ best\_score}{
     best\_score = loss\;
     noise = noise\_size\;
     best\_GAN = GAN\;
    }
   }
  }
 }
}

\Return{best\_GAN}
\caption{Pseudocode of the GAN training process}
\label{alg:GAN}
\end{algorithm}

\textbf{Step 2 : Attack generation and evaluation}

This step is to evaluate the current score of our IDS. To do so, we need to generate attacks with the GAN and gauge the robustness of our IDS against those attacks.

After the GAN has been trained at step 1, we use it to generate attacks. Generated attacks will use the functional features of attacks from the test set.

We evaluate the score of the IDS with those generated attacks. With previous notations, we consider that an instance $x$ is considered an actual attack by the IDS if $D(x) > 0.5$. The score is therefore the number of generated attacks $G(z)$ with $D(G(z)) > 0.5$, divided by the total number of generated attacks.

If the score has not improved in three rounds, we stop the algorithm.

\textbf{Step 3 : Search-based method}

In this step, we run a metaheuristic algorithm in order to generate additional attacks to further improve our detection system.

As our Search-based method, we used an hybrid genetic local-search algorithm. Indeed, local search and the Genetic Algorithm both have their pros and cons. The Genetic Algorithm is rather slow to converge whereas the Local search could converge to local optima. We chose to combine the two with an hybrid genetic algorithm~\cite{Hybrid}, as it has been demonstrated to have been more efficient in complex problems such as the Traveling Salesman~\cite{Salesman}.\par

The hybrid algorithm that we chose is a modification of the genetic algorithm: before proceeding to the selection process of the algorithm, every solution from the solution pool is improved by the local-search algorithm. As each solution is enhanced before the selection process, this algorithm allows for overall better performances, and usually a faster convergence than the standard Genetic Algorithm.

The goal of this metaheuristic algorithm is also to generate attacks against the IDS. Similarly to the proceedings of the GAN, functional features of our generated attacks will be from real attacks from the original dataset.

We first create a population of random solutions. We chose a population size of 30, as the recommended values in the literature are within the range of 30 to 80~\cite{popsize}. Having a bigger population affected the performances of our algorithm. 

Before the selection process, we optimize each solution of the population with a local search algorithm. The pseudocode for this local search method is given in algorithm \ref{alg:localsearch}.

\SetAlFnt{\small}
\begin{algorithm}[ht]
\SetAlgoLined

\SetKwProg{Fn}{Function}{ is}{end}\Fn{score(sol: array) : float}{
score = 1 - max(discriminator(sol),classifier(sol))\;
return score\;
}

\KwIn{Population of solutions (solutions), Discriminator, Classifier, functional features (func\_feat)}
\KwOut{Optimized population of solutions}
\BlankLine
\For{sol in solutions}{
\tcc{For each solution in the population, we slightly modify all the non functional characteristics to find the best solution in the neighborhood}
 \For{characteristic in sol}{
 \BlankLine
  \If{characteristic not in func\_feat}{
   modif = -0.01\;
   current\_value = characteristic\;
   best\_value = characteristic\;
   best\_score = score(sol)\;
   
   \tcc{We test all modifications from -0.01 to 0.01}
   \While{modif $<$ 0.01}{
    modif = modif + 0.001\;
    characteristic = current\_value + modif \;
    score = score(sol)\;
    \BlankLine
    \If{score $>$ best\_score}{
     best\_value = characteristic\;
     best\_score = score\;
    }
   }
   characteristic = best\_value\;
  }
 }
}
\Return{solutions}
\label{alg:localsearch}
\caption{Pseudocode of the used local search algorithm}

\end{algorithm}

Crossover is made by selecting two parents in the solution pool. We select only members of the population with the highest score (meaning, the attacks the most able to fool the IDS). The offspring will have the first half of its features from its first parent, and the other half from its second parent.

The mutation process is carried out to the entire population of children of this iteration. For each child, a non-functional feature selected at random is modified. The modification follows a uniform distribution, varying from -0.01 to 0.01.

Then, the new generation is equally composed of parents from the previous generation, and of its offspring. The fact of having members of the previous generation prevents the deterioration of the ability of the overall population to deceive the Intrusion Detection System.

We stop the hybrid genetic algorithm after 500 generations, or after 50 generations without improvement. These numbers were found to be experimentally sufficient for successfully training the four different IDS.

This population-based approach makes the solution pool iteratively evolve to better evade the detection system, and therefore generates a wide variety of adversarial attacks.

\textbf{Step 4 : IDS training}

In this final step, we aim to retrain the detection system for it to take the generated attacks into account. We train the IDS with :
\begin{itemize}
    \item All the attacks generated by the hybrid algorithm during its run at step 3.
    \item The trained GAN generated attacks from the training set.
    \item Examples from the original training set.
\end{itemize}


\subsection{Execution of SIGMA}
We executed SIGMA 
on the CalculQuebec Cloud service with the following computing resource : 15 X Intel Xeon @2,5Ghz, 128Go RAM, 10 core, 8 X Nvidia K20-GK110 GPU.  \par

The Pytorch module was used to implement all the neural networks.

In Table \ref{table:params}, we present all the parameters used to train the Neural Networks.

\begin{center}
\begin{table}[ht]
\caption{Training parameters for our Generative Adversarial Network, and for Neural Networks used as IDS.
}
\begin{tabular}{ |l|l|} 
\hline
Number of training epochs & 30 \\
Batch size & 64 \\
Learning rate & 0.01 \\
Loss function & $L1(\begin{bmatrix}
           u_{1} \\
           u_{2} \\
           \vdots \\
           u_{m}
         \end{bmatrix}
         , \begin{bmatrix}
           v_{1} \\
           v_{2} \\
           \vdots \\
           v_{m}
         \end{bmatrix})$ 
         $= \sum_{i} |u_i - v_i|$ \\
Optimizer & Adam \\
\hline
\end{tabular}
\label{table:params}

\end{table}
\end{center}

\subsection{Research questions}

To evaluate the effectiveness of SIGMA at improving the effectiveness of IDSs, we formulate the following two research questions:
\begin{itemize}
\item (RQ1) \emph{To what extent SIGMA can generate adversarial attacks able to deceive trained classifiers, acting as Intrusion Detection System?}\\
This research question aims to assess the effectiveness of
SIGMA at generating meaningful adversarial attack queries.
\item (RQ2) \emph{To what extent is the effectiveness of IDS improved using SIGMA?}\\
This research question aims to examine if through the successive re-training steps of SIGMA, IDSes are successfully improved. 
\end{itemize}

In the following, we describe the approach followed to answer RQ1, RQ2.

For RQ1, we use four different classification algorithms as IDS : Neural Network, Random Forest, Support Vector Machine and a Naive Bayes Classifier. 
We generate attacks against each of the IDS for all four attacks datasets (DOS, DDOS, Bruteforce, Infiltration) by using a GAN, trained with the methodology described above.

We compute the score of each of the detection systems for the GAN generated attacks, and therefore assess if SIGMA is able to deceive standard classification algorithms acting as IDS. \newline

For RQ2, we use a more complex intrusion detection system. We build an IDS consisting of two classifiers : an attack classifier, and a discriminator.
The attack classifier is trained with the entries from the original dataset, whereas the discriminator is trained with both regular attacks from the dataset as well as with generated entries to classify the input as a generated attack or as regular traffic.
Traffic is first analyzed by the discriminator to determine whether it is an adversarial instance or real traffic. If the input is labeled as real traffic, it then comes through the attack classifier whose role is to recognize attacks.
This architecture prevents from training the classifier with the adversarial examples, which could lead to a loss of performance for previously seen regular attacks because of overfitting to adversarial instances. It consists of a simple adaptation of the GAN discriminator to detect both generated instances and attacks from the dataset.
Therefore, since the goal of the discriminator is to identify generated instances, it will be the part of the IDS trained with the SIGMA generated attacks.

\begin{figure}[ht]
    \begin{center}
        \includegraphics[width=1\linewidth]{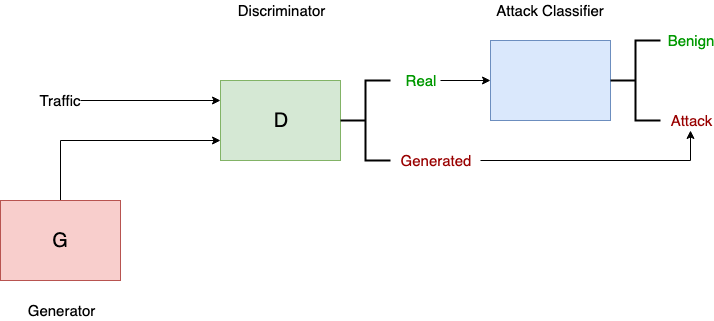}
    \end{center}
	
  	\caption{Diagram of the Intrusion detection system. Characteristics from the traffic are the inputs of the Discriminator, and goes into the attack Classifier if the Discriminator labels it as a real entry of the dataset.}
  	\label{fig:expSys}
\end{figure}

The overall process of the Intrusion Detection System studied is detailed on Fig. \ref{fig:expSys}

As attack classifier and discriminator, we used the same algorithms as for RQ1 : Neural Network, Random Forest, Support Vector Machine (SVM), and a Naive Bayes Classifier.

We study the largest dataset of the four (the DOS attacks dataset). For each round of improvement of SIGMA, we compute the score of the IDS.

To measure the performance of SIGMA, we compare our strengthened model to a baseline, in which the discriminator is trained only with GAN generated instances. We also verify that metaheuristics alone are not enough to train our system against generated adversarial attacks by comparing the model strengthened by SIGMA with a model trained only with the metaheuristics attacks, and submitting it to GAN generated attacks.

We will judge the quality of the reinforcement by : 

\begin{itemize}
    \item The speed of convergence of the detection rate.
    \item The value of the limit of the detection rate.
    \item The overall performance of the model for all iterations.
\end{itemize}


\begin{table*}[t]
\caption{Detection rates for test attacks from the dataset and GAN generated instances.}
        \centering
        \begin{tabular}{c|c|c|c|c|c|c|c|c}
        
         Classifier type & \multicolumn{2}{c|}{Neural Net} &\multicolumn{2}{c|}{Random Forest}& \multicolumn{2}{c|}{SVM} & \multicolumn{2}{c}{Naive Bayes} \\
        \hline
        Type of attack & Normal & Adversarial & Normal & Adversarial & Normal & Adversarial & Normal & Adversarial \\
         \hline
        DOS & 94,9\% & 0\% & 99,8\% & 0\% & 97,6\% & 0\% & 95,9\% & 0\% \\
        DDOS & 98,6\% & 29,9\% & 99,9\% & 0\% & 98,3\% & 47,1\% & 97,1\% & 0\% \\
        Bruteforce & 95,6\% & 1,9\% & 99,9\% & 0\% & 96,3\% & 0\% & 97,8\% & 0\% \\
        Infiltration & 95,4\% & 5,8\% & 100\% & 0\% & 96,2\% & 4,1\% & 97,4\% & 0\% \\
         
        \end{tabular}
        \label{table:DRGAN}

\end{table*}

It should be noted that for the first iteration of the algorithm, the discriminator has not yet been trained : the generator is thus only trained against the classifier at the first iteration.

\subsection{ Results of the Evaluation of SIGMA}\label{results}
In this section we present the answers to our two research questions that aim to evaluate SIGMA.\\
\textbf{\emph{RQ1: To what extent SIGMA can generate adversarial attacks able to deceive trained classifiers, acting as Intrusion Detection System?}}

\begin{table*}[t]
\caption{Evolution of the detection rates of adversarial attacks for our model}
        \centering
        \begin{tabular}{c|c|c|c|c|c|c|c|c}
        
         Classifier type & \multicolumn{2}{c|}{Neural Net} &\multicolumn{2}{c|}{Random Forest}& \multicolumn{2}{c|}{SVM} & \multicolumn{2}{c}{Naive Bayes} \\
        \hline
        Iteration number & Normal & Reinforced & Normal & Reinforced & Normal & Reinforced & Normal & Reinforced \\
         \hline
        1 & 0\% & 0\% & 0\% & 0\% & 0\% & 0\% & 0\% & 0\% \\
        2 & 6,3\% & 0\% & 51\% & 0\% & 100\% & 100\% & 100\% & 100\% \\
        3 & 49\% & 100\% & 99\% & 100\% & 100\% & 100\% & 100\% & 100\% \\
        4 & 100\% & 100\% & 18\% & 100\% & 100\% & 100\% & 100\% & 100\% \\
        5 & 68\% & 100\% & 100\% & 100\% & 100\% & 100\% & 100\% & 100\% \\
        6 & 100\% & 100\% & 100\% & 100\% & 100\% & 100\% & 100\% & 100\% \\
        7 & 100\% & 100\% & 100\% & 100\% & 100\% & 100\% & 100\% & 100\% \\
         
        \end{tabular}
        \label{table:SIGMA}
        
\end{table*}

\begin{figure}[ht]
    \begin{center}
        \hspace*{-0.05in}
        \includegraphics[width=1.13\linewidth]{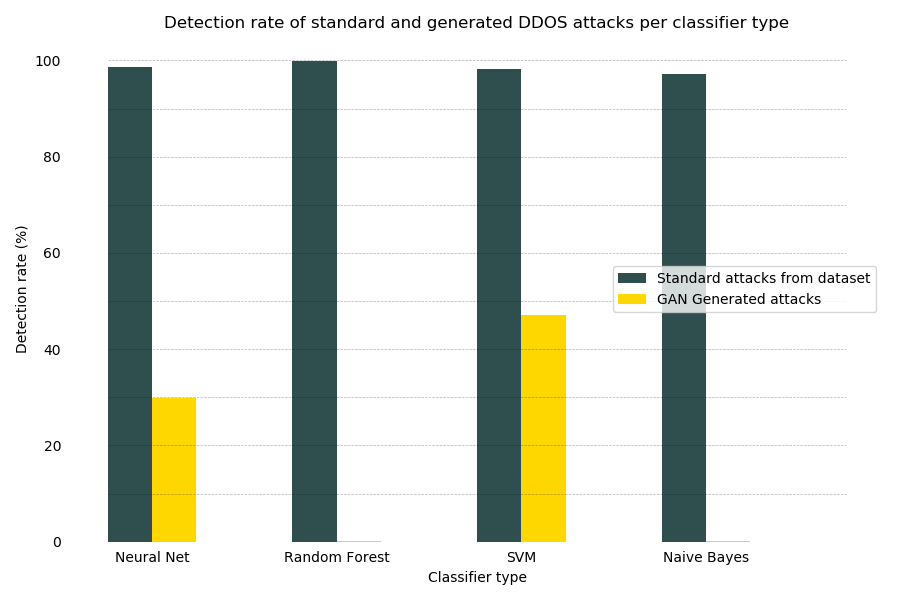}
    \end{center}
	
  	\caption{Detection scores per classifier with test set attacks and generated attacks for the DDOS dataset.}
  	\label{fig:barchartDDOS}
\end{figure}

The results of the detection of normal and generated attacks are presented in Table \ref{table:DRGAN}, and on Fig. \ref{fig:barchartDDOS}. \par
All four classifiers in our study (Neural Network, Random Forest, SVM, Naive Bayes) have good results in classifying standard entries of the datasets. Even though our classifiers are standard machine learning algorithms, they are  sufficient to obtain high accuracy, with the Random Forest algorithm performing with the best results with an overall 99,9\% accuracy, followed by the Support Vector Machine with 97,1\%. In fact, those two algorithms have often been used in intrusion detection thanks to their good performances~\cite{SVMRFIDS}.

However, the generated attacks detection rates is significantly low for all classifiers with most type of attacks. Both the Random Forest and the Naive Bayes are utterly unable to detect the GAN generated adversarial attacks. The neural network and the SVM are the most resilient classifiers, but the generator still manages to deceive our IDS with over a 90\% evasion rate for the DOS, Bruteforce and Infiltration attacks.

The results show very good performance of the Generative Adversarial Network for all different types of attacks. It is therefore possible to generate attacks able to fool Machine Learning based classifiers for all four types of attacks.

\begin{figure*}[ht]
    \begin{center}
        \includegraphics[width=1\linewidth]{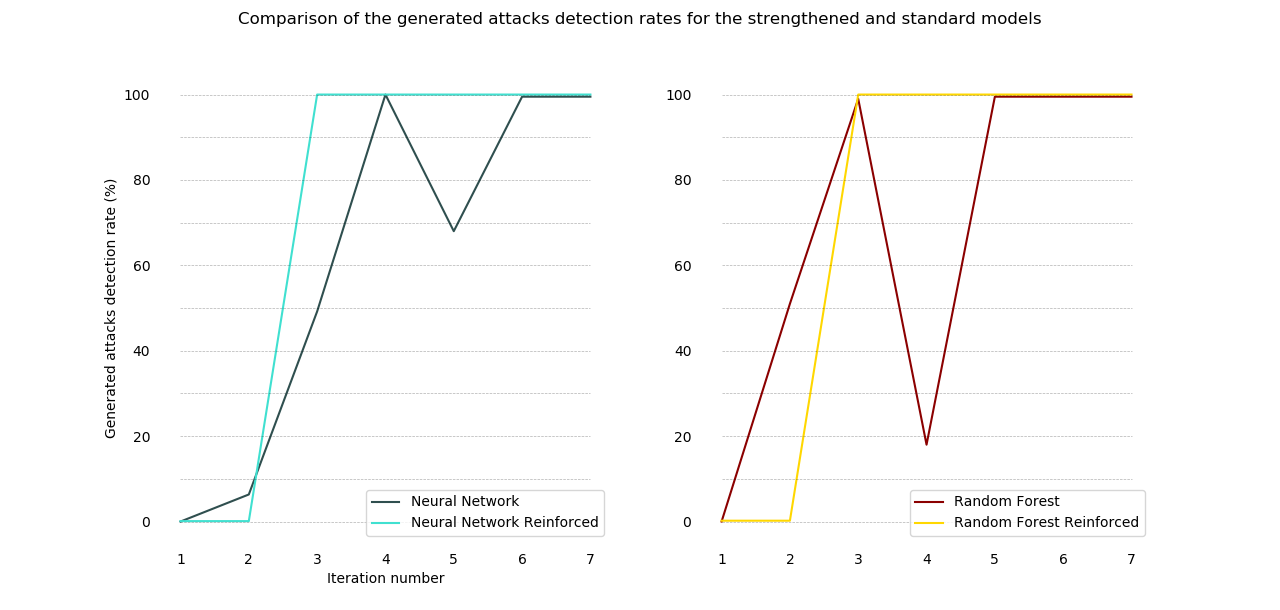}
    \end{center}
	
  	\caption{Time evolution of our reinforced model with two different classifiers (Neural Network and Random Forest) and their corresponding standard model.}
  	\label{fig:barcharttensteps}
\end{figure*}

\textbf{\emph{RQ2: To what extent is the effectiveness of IDS improved using SIGMA?}}\\
We compared the evolution of our model trained with the hybrid local-search-genetic reinforcement and adversarial attacks with a model trained only with adversarial attacks. 

The results are presented in Table \ref{table:SIGMA} and Fig. \ref{fig:barcharttensteps}.

First, we notice that both models with the SVM and the Naive Bayes as classifiers only need one step to detect adversarial attacks: those two classifiers are the most able to generalize from the previously seen data. The generated attacks detection rate converges after only one iteration for both the strengthened and the standard model.\par

The multi-layer Neural Network and the Random Forest standard models both take time to converge to a 100\% generated attacks detection rate: 6 iterations for the model with the Neural Network as classifier, 5 iterations for the Random Forest model. 
Furthermore, we also note that both models suffered from overfitting: their performance increased (until iteration 4 and 3 respectively) before dropping significantly by 32\% and 81\%.

The SIGMA method improved the models' results: as we can see, the strengthened model converged faster than the standard model to a 100\% detection rate for both the Neural Network and the Random Forest classifiers; the reinforced versions took only two iterations to detect all adversarial instances, that is to say respectively four and three iterations less. As the other two classifiers, namely the SVM and the Naive Bayes classifier, detected all attacks from iteration 2, the reinforcement method did not affect their performance. \par

Furthermore, we can observe that the SIGMA method prevented the Neural Network and the Random Forest model from overfitting to generated attacks, therefore preventing a performance drop of the algorithm. The combination of the metaheuristic algorithm and the Generative Adversarial Network permitted to generate a sufficiently wide variety of attacks; avoiding fitting closely to 
previously seen attacks. 
\begin{center}
\begin{table}[ht]
\caption{Evolution of the detection rate of adversarial attacks for a model trained only with Metaheuristics generated attacks.}
\begin{tabular}{|c|c|c|c|c|} 
\hline

Iteration & Neural Net & Random Forest & SVM & Naive Bayes \\
\hline
1 & 0 & 0 & 0 & 0\\
2 & 0 & 0 & 0 & 0\\
3 & 0 & 0 & 40.1 & 0\\
4 & 0 & 0 & 100 & 0\\
5 & 0 & 0 & 58.1 & 0\\
6 & 0 & 0 & 0.8 & 0\\
7 & 0 & 0 & 52.9 & 0\\
\hline
\end{tabular}
\label{table:metaonly}

\end{table}
\end{center}

Table \ref{table:metaonly} presents the results of models trained only from the Metaheuristics generated attacks. From these results, we can also conclude that Metaheuristics alone are not sufficient to train an IDS against generated adversarial attacks: every classifier, except the Support Vector Machine, was utterly unable to detect any instance generated by our Wasserstein GAN. The SVM stands out from the other classifiers thanks to its ability to generalize, but fails at consistently detecting GAN generated attacks. \par
We can therefore conclude that the attacks generated by the Metaheuristics algorithm complement the ones generated by the Generative Adversarial Networks, as the Metaheuristics algorithm alone was not enough to successfully train the IDS.

\section{Threats to validity}\label{threats}
This section discusses the threats to validity of our study following common guidelines for empirical studies \cite{Yin}

\emph{Construct validity} threats concern the relation between
theory and observation. This is mainly due to possible mistakes in the generation of attacks. Even though we kept the functional features of real attack untouched for our generated attacks, we can not guarantee that the generated attacks metrics are indeed plausible attacks. 

\emph{Internal validity} threats concern the selection of tools and analysis methods. We split the dataset into a training and a test set in order to ensure the validity of our results. This prevents having a biased evaluation of our model. 
As the aim of the method was to try to detect as many generated attacks as possible, we chose to study the generated attacks detection rate as a metric to gauge the quality of the strengthening. 

\emph{Reliability validity} threats concern the possibility to replicate our study. All the tools used in this study are open-source.

\emph{Conclusion validity} threats concern the relation between
treatment and the outcome. We paid attention to not make too broad statements about the performances of our model. 

\emph{External validity} threats concern the possibility to generalize our results. The results of the SIGMA method have to be interpreted carefully, as they may depend on the dataset used to run the experiment and on the used Intrusion Detection System.
The iterative strengthening method has only been studied for DOS attacks of the CICIDS 2017 dataset \cite{newdataset}. We used four different classifiers acting as IDS, and were able to significantly improve the results of two of the four IDS. 
We therefore suggest that our results can be generalized to other detection systems and other datasets.


\section{Implication for Practitioners and the Industry}\label{implications}

Artificial Intelligence is a really powerful tool that could and will be used in future cybersecurity systems : IBM's Watson is one of the illustrations of the application of Machine Learning in this field. Nonetheless, this work illustrated possible vulnerabilities of such systems as Artificial Intelligence can also be leveraged by attackers to disrupt detection systems.

Generative Adversarial Networks can be used to forge almost undetectable adversarial attacks for systems that have not already faced such attacks. Our method confronted our studied Intrusion Detection Systems with attacks generated with both GANs and Metaheuristics in order to improve the systems resilience, as our analysis has shown that the more attacks the system faces, the more it will be able to efficiently generalize to other potential attacks.

Repetitively training an IDS with generated attacks is a way to anticipate for every possible generative scheme that could target the system. By doing so, our method SIGMA is able to detect all the attacks generated by our GAN, thus preventing future intrusion by adversarial generated attacks.

Such methods should be applied to any AI-based cybersecurity system in the industry to preemptively confront them to new types attacks, therefore preventing them from such possible threats.


\section{Conclusion}
\label{conclusion}
The novel ability to use Machine Learning techniques to generate adversarial attacks requires the development of a robust IDS able to detect unusual behaviors. Generative Adversarial Networks are both a terrible weapon for detection systems, and an incredible opportunity to preemptively strengthening IDSs against adversarial attacks. \par
We have shown experimentally that it is possible to effectively evade intrusion detection classifiers with Generative adversarial networks. We demonstrated the possibility to forge undetected adversarial attacks with GANs against four standard Machine Learning algorithms acting as IDS, with the generated attacks detection rates dropping near 0\% for most of them. \par

To prevent adversarial generated attacks, we presented in this paper a method SIGMA, to improve the robustness of IDS. This method is based on the iterative generation of attacks by a Machine Learning Generative algorithm and Metaheuristics.
We have shown that applying this method to Machine Learning based IDS can speed up the convergence of the generated attacks detection rate, and prevent overfitting to previously seen generated attacks. \par
Our model may help design Intrusion detection systems robust against recurrent generative attacks and improve the security of Machine Learning systems. \par
Further considerations are the explorations of other more complex detection algorithms, such as Recurrent Neural Networks, the application of the SIGMA method to other datasets and the design of a distributed detection system robust to adversarial attacks.


%




\ifCLASSOPTIONcaptionsoff
  \newpage
\fi




\begin{thebibliography}{1}

  
\bibitem{IOTchallenges}
Javier~Lopez Rodrigo Roman Jianying~Zhou.
\newblock On the features and challenges of security and privacy in distributed Internet of things.
\newblock {\em Computer Networks}, 2013.
  
\bibitem{GAN}
Ian Goodfellow, Jean Pouget-Abadie, Mehdi Mirza, Bing Xu, David Warde-Farley,
  Sherjil Ozair, Aaron Courville, and Yoshua Bengio.
\newblock Generative adversarial nets.
\newblock In {\em Advances in neural information processing systems}, pages
  2672-2680, 2014.
  
\bibitem{Metaheuristic}
Jan, Sadeeq, et al. 
\newblock Automatic generation of tests to exploit XML injection vulnerabilities in web applications. 
\newblock IEEE Transactions on Software Engineering 45.4 : pages 335-362, 2017.

  
\bibitem{Mirai}
Constantinos Kolias, Georgios Kambourakis, Angelos Stavrou and Jeffrey Voas
\newblock DDoS in the IoT: Mirai and other botnets
\newblock In {\em Computer(50),7}, pages
  80-84, 2017.
  
\bibitem{MLforIDS}
Tsai, C., Hsu, Y., Lin, C. and Lin, W. . 
\newblock{Intrusion detection by machine learning: A review.} \newblock In {Expert Systems with Applications, 36(10), pages 11994-12000.}, 2009.

\bibitem{Yin}
R. K. Yin \newblock
Case Study Research: Design and Methods\newblock
Third Edition,3rd ed. SAGE Publications, 2002.
  
\bibitem{IDSGAN}
Z.~{Lin}, Y.~{Shi}, and Z.~{Xue}.
\newblock {IDSGAN: Generative Adversarial Networks for Attack Generation
  against Intrusion Detection}.
\newblock {\em arXiv e-prints}, 2018.

\bibitem{MalwareGAN}
Weiwei Hu and Ying Tan.
\newblock Generating adversarial malware examples for black-box attacks based on gan.
\newblock {\em arXiv preprint arXiv:1702.05983}, 2017.

\bibitem{disruptiveGAN}
Anish Athalye, Logan Engstrom, Andrew Ilyas and Kevin Kwok. \newblock Synthesizing robust adversarial examples. 
\newblock {\em arXiv preprint arXiv:1707.07397.}, 2017.

\bibitem{searchbased}
Cordy, M., Muller, S., Papadakis, M.,  Le Traon, Y. 
\newblock Search-based test and improvement of machine-learning-based anomaly detection systems. 
\newblock In Proceedings of the 28th ACM SIGSOFT International Symposium on Software Testing and Analysis, pages 158-168. ACM., 2019


\bibitem{KDD99}
H~G{\"u}nes Kayacik, A~Nur Zincir-Heywood, and Malcolm~I Heywood.
\newblock Selecting features for intrusion detection: A feature relevance
  analysis on kdd 99 intrusion detection datasets.
\newblock In {\em Proceedings of the third annual conference on privacy,
  security and trust}, 2005.
 
 
\bibitem{ELM}
S. Prabavathy, K. Sundarakantham and S.M. Shalinie 
\newblock Design of cognitive fog computing for intrusion detection in internet of things.
\newblock Journal of Communications and Networks, 20(3), pages 291-298, 2018.

\bibitem{NSLSVM}
Dhanabal, L.,  Shantharajah, S. P.
\newblock A study on NSL-KDD dataset for intrusion detection system based on classification algorithms. 
\newblock International Journal of Advanced Research in Computer and Communication Engineering, 4(6), pages 446-452, 2015.

\bibitem{HGAspeed}
Oh, I. S., Lee, J. S., Moon, B. R. \newblock 
Hybrid genetic algorithms for feature selection. 
\newblock IEEE Transactions on pattern analysis and machine intelligence, 26(11), pages 1424-1437, 2004.

\bibitem{photorealistGAN}
Christian Ledig, Lucas Theis, Ferenc Husz{\'a}r, Jose Caballero, Andrew
  Cunningham, Alejandro Acosta, Andrew Aitken, Alykhan Tejani, Johannes Totz,
  Zehan Wang, et~al.
\newblock Photo-realistic single image super-resolution using a generative
  adversarial network.
\newblock {\em arXiv preprint}, 2017.

\bibitem{KDDcritique}
John McHugh.
\newblock Testing intrusion detection systems: a critique of the 1998 and 1999
  darpa intrusion detection system evaluations as performed by lincoln
  laboratory.
\newblock {\em ACM Transactions on Information and System Security (TISSEC)},
  3(4) pages 262-294, 2000.

\bibitem{DistributedGAN}
Aidin Ferdowsi and Walid Saad. 
\newblock Generative Adversarial Networks for Distributed Intrusion Detection in the Internet of Things. 
\newblock arXiv preprint arXiv:1906.00567, 2019
  
\bibitem{benchmark}
Iman Sharafaldin, Amirhossein Gharib, Arash~Habibi Lashkari, and Ali~A
  Ghorbani.
\newblock Towards a reliable intrusion detection benchmark dataset.
\newblock {\em Software Networking}, pages 177-200, 2018.

\bibitem{SGA}
Mitchell, Melanie, John H. Holland, and Stephanie Forrest. \newblock "When will a genetic algorithm outperform hill climbing." 
\newblock Advances in neural information processing systems. 1994.

\bibitem{Salesman}
Ulder, Nico LJ, et al. 
\newblock Genetic local search algorithms for the traveling salesman problem.
\newblock International Conference on Parallel Problem Solving from Nature. Springer, Berlin, Heidelberg, 1990.

\bibitem{Hybrid}
Ishibuchi, Hisao, and Tadahiko Murata. 
\newblock Multi-objective genetic local search algorithm.
\newblock Proceedings of IEEE international conference on evolutionary computation. IEEE, 1996.


\bibitem{newdataset}
Iman Sharafaldin, Arash~Habibi Lashkari, and Ali~A Ghorbani.
\newblock Toward generating a new intrusion detection dataset and intrusion
  traffic characterization.
\newblock In {\em ICISSP}, pages 108-116, 2018.


\bibitem{RNNIDS}
George Loukas, Tuan Vuong, Ryan Heartfield, Georgia Sakellari, Yongpil Yoon and Diane Gan
\newblock Cloud-based cyber-physical intrusion detection for vehicles using deep learning. 
\newblock Ieee Access, 6, pages 3491-3508, 2017

\bibitem{Genetic}
Gen, Mitsuo, and Lin Lin. 
\newblock Genetic Algorithms. 
\newblock Wiley Encyclopedia of Computer Science and Engineering : 1-15, 2007

\bibitem{Wasserstein}
Arjovsky, Martin, Soumith Chintala, and Léon Bottou. \newblock Wasserstein GAN 
\newblock arXiv preprint arXiv:1701.07875, 2017.

\bibitem{popsize}
Cobb, H. G., Grefenstette, J. J. \newblock
Genetic algorithms for tracking changing environments. \newblock
Naval Research Lab Washington DC, 1993.

\bibitem{AIsigns}
Eykholt, K., Evtimov, I., Fernandes, E., Li, B., Rahmati, A., Xiao, C., ...  Song, D. 
\newblock Robust physical-world attacks on deep learning visual classification. 
\newblock In Proceedings of the IEEE Conference on Computer Vision and Pattern Recognition (pages 1625-1634), 2018

\bibitem{SVMRFIDS}
Hasan, Md Al Mehedi, et al.
\newblock Support vector machine and random forest modeling for intrusion detection system (IDS). 
\newblock Journal of Intelligent Learning Systems and Applications 6.01: 45, 2014

\end{thebibliography}
%

%

\begin{IEEEbiographynophoto}{Simon Msika}
received a M.S. degree from Ecole polytechnique, Paris, France, before pursuing a M.S. in Cybersecurity at Ecole Polytechnique de Montreal, Canada in 2018. His research interests include Machine learning applied to security, and Cryptography.
\end{IEEEbiographynophoto}

\begin{IEEEbiographynophoto}{Alejandro Quintero}
received the engineer’s degree in computer engineering from Los Andes,Colombia, in 1983. In June 1989 and in 1993,he received a diploma of advanced studies and a Doctorate in computer engineering, respectively, from the INPG Grenoble and Universite Joseph Fourier, Grenoble, France. He is currently a full professor  at  the  Department  of  Computer  Engineering  of Ecole Polytechnique de Montreal, Canada. His main research interests include services  and  applications  related  to  mobile  computing, network security, networks infrastructures and next generation mobile  networks.  He  is  the  co-author  of  one  book,  as  well  as  more than 60 other technical publications including journal and proceedings papers.
\end{IEEEbiographynophoto}


\begin{IEEEbiographynophoto}{Foutse  Khomh}
is an associate professor at Polytechnique Montréal and FRQ-IVADO Research Chair on Software Quality Assurance for Machine Learning Applications. He received a Ph.D in Software Engineering from the University of Montreal in 2010, with the Award of Excellence. His research interests include software maintenance and evolution, engineering of cyber-physical systems, empirical software engineering, machine learning systems engineering, and software analytic. His work has received three ten-year Most Influential Paper (MIP) Awards, and five Best/Distinguished paper Awards. He serves on the program committees, steering committees, and editorial boards of several international conferences and top international journals in software engineering.
\end{IEEEbiographynophoto}




\end{document}